\documentclass[11pt,a4paper]{article}

\usepackage[truedimen,margin=34mm]{geometry} 

\usepackage{mathrsfs}
\usepackage{amssymb}
\usepackage{amsmath}
\usepackage{ascmac}
\usepackage{amsthm}
\usepackage[dvipdfmx]{graphicx}

\usepackage{color}

\usepackage{titlesec}
\titleformat*{\section}{\large\bfseries}
\titleformat*{\subsection}{\it}

\newtheorem{algo}{Algorithm}

\def\al{{\alpha}}

\def\ep{{\varepsilon}}

\def\si{{\sigma}}

\def\bal{{\text{\boldmath $\alpha$}}}
\def\bbe{{\text{\boldmath $\beta$}}}
\def\bga{{\text{\boldmath $\gamma$}}}
\def\bpi{{\text{\boldmath $\pi$}}}

\def\bth{{\text{\boldmath $\theta$}}}
\def\bphi{{\text{\boldmath $\phi$}}}

\def\balh{{\widehat \bal}}

\def\sih{{\widehat \si}}

\def\bbeh{{\widehat \bbe}}

\def\bphih{{\widehat \bphi}}

\def\bthh{{\widehat \bth}}

\def\pt{\widetilde{p}}

\def\at{\widetilde{a}}
\def\bat{\widetilde{\a}}
\def\bpt{\widetilde{\p}}

\def\1{{\text{\boldmath $1$}}}
\def\a{{\text{\boldmath $a$}}}
\def\w{{\text{\boldmath $w$}}}
\def\x{{\text{\boldmath $x$}}}
\def\y{{\text{\boldmath $y$}}}
\def\z{{\text{\boldmath $z$}}}

\def\p{{\text{\boldmath $p$}}}

\def\barx{\bar{\x}}

\def\E{{\rm E}}

\def\Dir{\rm Dir}

\def\mit{\vspace{-0.23cm}\item}

\title{{\bf Latent Mixture Modeling for Clustered Data}\footnote{This version: \today}}

\date{}

\begin{document}

\maketitle

\vspace{-1cm}\noindent
SHONOSUKE SUGASAWA\\
{\it Risk Analysis Research Center, The Institute of Statistical Mathematics}\\
GENYA KOBAYASHI\\
{\it Graduate School of Social Sciences, Chiba University}\\
YUKI KAWAKUBO\\
{\it Graduate School of Social Sciences, Chiba University}

\vspace{0.8cm}\noindent
{\large\bf Abstract.}\ \ \ 
This article proposes a mixture modeling approach to estimating cluster-wise conditional distributions in clustered (grouped) data.  
We adapt the mixture-of-experts model to the latent distributions, and propose a model in which each cluster-wise density is represented as a mixture of latent experts with cluster-wise mixing proportions distributed as Dirichlet distribution.
The model parameters are estimated by maximizing the marginal likelihood function using a newly developed Monte Carlo Expectation-Maximization algorithm. 
We also extend the model such that the distribution of cluster-wise mixing proportions depends on some cluster-level covariates.
The finite sample performance of the proposed model is compared with some existing mixture modeling approaches as well as linear mixed model through the simulation studies. 
The proposed model is also illustrated with the posted land price data in Japan.


\section{Introduction}
Grouped or clustered data often arise in many scientific fields such as econometrics, epidemiology, and genetics. 
Although the mixed-effects model (Demidenko, 2004) has been widely used for such data, it fundamentally aims at modeling conditional means in each cluster, which could be inappropriate if the data distribution is skewed or multimodal.
As an alternative modeling strategy, the finite mixture model (McLachlan and Peel, 2000) has been extensively applied for its flexibility to capture the within-cluster heterogeneity in the data.
For modeling independent data, the mixture model with covariates was originally proposed in Jacob et al. (1991), known as mixture-of-experts.
To date, a large body of literature has been concerned with flexible modeling of the conditional density for independent data. 
For example, see Jordan and Jacobs (1994), Hurn et al. (2003), Geweke and Keane (2007), Villani et al. (2009), Villani et al. (2012) and Nguyen and McLachlan (2016).

However, the existing models for independent data are not suitable for estimating cluster-wise conditional distributions.
If we globally apply the mixture models to a whole dataset ignoring the clustering labels (we call global mixture modeling), the estimated conditional distributions are the same over all clusters, which is clearly inappropriate in clustered data analysis. 
On the other hand, applying the mixture models independently to each cluster in order to capture the cluster heterogeneity (we call local mixture modeling) leads to unstable results since the within-cluster samples sizes are usually not large in practice.  
Hence, another flexible modeling strategy for clustered data is desired.
Up to now, several methods have been proposed for modeling cluster-wise distributions.
Rubin and Wu (1997) proposed a mixture of linear mixed-effects models.
Sun et al. (2007) developed a mixture of linear models with the random effects used in the generalized linear model for the mixing proportions.
Rosen et al. (2000) and Tang and Qu (2016) used the generalized estimating equation approach to estimate the component distributions by incorporating the correlations within clusters.

In this article, we propose a compromised model between the global and local mixture modeling.
Note that the local mixture model can be expressed as 
$$
f_i(y|\x)=\sum_{k=1}^K \pi_{ik}h_{ik}(y|\x),
$$
where $y$ is the response variable, $\x$ is the vector of covariates, and $h_{ik}$ is the component distribution for the $k$th component of the $i$th cluster with the mixing proportion $\pi_{ik}$ satisfying $\sum_{k=1}^K\pi_{ik}=1$. 
Since the within-cluster sample size is usually small in practice, $h_{ik}(y|\x)$ would not be stably estimated.
Hence, we restrict $h_{ik}(y|\x)=h_{k}(y|\x)$, that is, the component distributions are the same over all the clusters like global modeling.  
Then the model reduces to 
$$
f_i(y|\x)=\sum_{k=1}^K \pi_{ik}h_{k}(y|\x),
$$
which can be interpreted as there exists $K$ latent distributions and each cluster-wise distribution $f_i(y|\x)$ is expressed by these distributions with cluster-wise mixing proportions $\pi_{ik}$.
Hence, as long as $K$ is a moderate number, one can estimate $K$ component distributions with reasonable accuracy.
On the other hand, estimating unstructured $\pi_{ik}$ is not feasible since the number of $\pi_{ik}$'s grows as the number of clusters increases.   
To overcome this difficulty, we assume that the vector of proportions $\bpi_i=(\pi_{i1},\ldots,\pi_{iK})^t$ that characterizes the conditional distribution of the $i$th cluster, is a realization from a multivariate distribution.
Therefore, $\bpi_i$ plays a similar role to the random effect in the context of the mixed-effects model.
As a distribution of $\bpi_i$, we use the Dirichlet distribution, which allows us to develop a tractable estimating method for model parameters.

In this article, the model parameters are estimated based on a likelihood-based approach. 
The model can be viewed as a three-stage hierarchical model, where the first stage consists of the model for the response variable, the second stage consists of the latent variables which assign the latent distribution, and the third stage consists of the model for the mixing proportions. 
We develop a Monte Carlo Expectation-Maximization (MCEM) algorithm (Dempster et al., 1977; Wei and Tanner, 1990) for parameter estimation of which the E-step is consist of a simple Gibbs sampling scheme for imputing the latent variables. 
Since the number of latent distributions $K$ is generally unknown, we consider selecting $K$ based on  the Akaike information criteria (AIC) or Bayesian information criteria (BIC), where the maximum log-marginal likelihood can be easily computed from a simple Monte Carlo approximation.

The rest of the paper is organized as follows: 
Section \ref{sec:model} describes the proposed model in detail and develops the MCEM algorithm for maximizing the marginal likelihood. 
In Section \ref{sec:num}, the performance of the proposed method is demonstrated along with some existing methods through simulation studies.
An application to the real data set is also presented. 
In Section \ref{sec:diss}, some discussion is provided.

\section{Latent Mixture Model}\label{sec:model}


\subsection{Model setup}
Suppose that we have the clustered (grouped) observations $y_{ij}$, $i=1,\dots,m$, $j=1,\dots,n_i$, with an associated $p$-dimensional vector of covariates $\x_{ij}$.
Let $f_i(y|\x)$ be a density or probability mass function of $y_{ij}$ given $\x_{ij}$, which are the same within clusters but different across clusters. 
Our aim is to estimate the cluster-wise conditional density $f_i(y|\x)$ from the data set $\{y_{ij},\x_{ij}\}$.
To this end, we consider the following latent mixture model:
\begin{equation}\label{model}
f_i(y|\bpi_i,\x,\bphi)=\sum_{k=1}^K \pi_{ik}h_k(y|\x,\bphi_k),
\end{equation}
where $\pi_{ik}$ is the weight for the $k$th component in the $i$th cluster, $h_k(\cdot|\cdot,\bphi_k), \ k=1,\ldots,K$ are the latent conditional densities characterized by the parameter $\bphi_k$, and $K$ is the unknown number of latent densities.
Moreover, we assume that the mixing proportions $\bpi_i$'s are independent realizations from the Dirichlet distribution with the density
\begin{equation}\label{Dir}
p(\bpi_i|\bal)=\frac{\Gamma\big(\sum_{k=1}^K\al_k\big)}{\prod_{k=1}^K\Gamma(\al_k)}\prod_{k=1}^K\pi_{ik}^{\al_k-1}
\end{equation}
for $i=1,\ldots,m$, where $\Gamma(\cdot)$ denotes the gamma function and $\bal=(\al_1,\ldots,\al_K)^t$ is a vector of unknown parameters.
In this article, we let \eqref{model} and \eqref{Dir} together denote the latent mixture model. 
The unknown model parameters to be estimated are $\bphi_1,\ldots,\bphi_K$ in latent distributions and $\bal$ in the Dirichlet distribution. 
Under the setting (\ref{model}) and (\ref{Dir}), taking expectation of $\pi_{ik}$ with respect to Dir$(\bal)$, we have
\begin{equation}\label{mar}
f_i(y|\x,\bal,\bphi)=\sum_{k=1}^K p_kh_k(y|\x,\bphi_k), \ \ \ \ \ 
p_{k}=\frac{\alpha_{k}}{\sum_{\ell=1}^K\alpha_{\ell}},
\end{equation}
which is referred to the marginal model, and is common over all the clusters.
Hence, we can observe that $\bpi_i$ characterizes the cluster-wise conditional density and plays a similar role to the random effects in the context of mixed-effects models. 
The mixing proportion $\bpi_i$ can be estimated by the conditional expectation $\E[\bpi_i|Y]$, where $Y$ is a set of all the response variables.
Under (\ref{model}) and (\ref{Dir}), response variables in different clusters are mutually independent, so that it holds $\E[\bpi_i|Y]=\E[\bpi_i|Y_i]$ with $Y_i=\{y_{i1},\ldots,y_{in_i}\}$.
Then, if the model parameters are known, the estimator of the cluster-wise conditional density is given by
\begin{equation}\label{Est}
\tilde{f}_i(y|\x,\bal,\bphi)=\sum_{k=1}^K \E[\pi_{ik}|Y_i]h_k(y|\x,\bphi_k).
\end{equation}
Generally speaking, the conditional expectation $\E[\pi_{ik}|Y_i]$ tends close to the marginal mean $p_k$ if the cluster-specific sample size $n_i$ is small, so that the estimated conditional density would be close to the marginal model (\ref{mar}).
On the other hand, in clusters with relatively large $n_i$, the estimated conditional density might vary from the marginal model (\ref{mar}), depending on the information of $Y_i$.
Therefore, this model allows us to carry out a kind of shrinkage estimation of the cluster-wise conditional densities.

As often done in estimating mixture models, by introducing the latent component indicator $z_{ij}\in\{1,\dots,K\}$, the proposed model (\ref{model}) and (\ref{Dir}) can be expressed in the three-stage hierarchical model:
\begin{equation}\label{hie}
\begin{split}
\text{1st stage:} \ \ \ &y_{ij}|\x_{ij},(z_{ij}=k) \sim F_k(\x_{ij},\bphi_k),\\
\text{2nd stage:} \ \ \ &z_{ij}|\bpi_i\sim \text{Cat}(K,\bpi_i), \\
\text{3rd stage:} \ \ \ &\bpi_i\sim \Dir(\bal),
\end{split}
\end{equation}
where $F_k$ is the distribution having density $h_k$, and $\text{Cat}(K,\bpi_i)$ is the categorical distribution on $\{1,\ldots,K\}$ with the probability vector $\bpi_i$.
In hierarchy (\ref{hie}), $\z_{ij}$ and $\bpi_i$ are the latent variables.
The latent density $h_k$ is determined by the user and the generalized linear model is an attractive choice.
For example, $F_k(\x_{ij},\bphi_k)= N(\x_{ij}^t\bbe_k,\sigma_k^2)$ when $y_{ij}$ is a continuous variable, and $F_k(\x_{ij},\bphi_k)=\text{Po}(\exp(\x^t\bbe_k))$ when $y_{ij}$ is a counting variable.


\subsection{Monte Carlo EM algorithm for parameter estimation}\label{sec:est}
For completion of the conditional density (\ref{Est}), we need to estimate the unknown model parameters $\bth=\{\bphi_1,\ldots,\bphi_K, \bal\}$ based on the data. 
Under the hierarchical formulation \eqref{hie}, the marginal likelihood function $L(\bth)$ is expressed as 
\begin{align*}
L(\bth)=\left(\frac{\Gamma(\sum_{k=1}^K\al_k)}{\prod_{k=1}^K \Gamma(\al_k)}
\right)^m
\prod_{i=1}^m\sum_{\z_i}
 \frac{ \prod_{k=1}^K \Gamma(\sum_{j=1}^{n_i}w_{ijk} + \al_k) }{ \Gamma(n_i + \sum_{k=1}^K\al_k) }
\left( \prod_{j=1}^{n_i}\prod_{k=1}^K h_k(y_{ij}|\x_{ij},\bphi_k)^{w_{ijk}} \right),
\end{align*}
where $w_{ijk}=I(z_{ij}=k)$ and $\sum_{\z_i}$ denotes the summation over the all combination of $\z_i\in \{1,\ldots,K\}^{n_i}$.
Hence, a direct maximization of the marginal likelihood is not feasible since evaluation of the likelihood function $L(\bth)$ requires the summation over $K^{n_i}$ elements for each $i$, which is computationally prohibitive even for small $K$.  
Moreover, since the functional form of $L(\bth)$ is complex and not familiar, the brute force maximization of $L(\bth)$ is not realistic.

Instead, we exploit the hierarchical representation (\ref{hie}) and develop the EM algorithm (Dempster et al., 1977) which indirectly and iteratively maximizes $L(\bth)$.
Let $\bpi=\left\{\bpi_1,\ldots,\bpi_m\right\}$ and $\z=\left\{\z_1,\ldots,\z_m\right\}$.
Then, the complete log-likelihood function $\ell^c$ of (\ref{hie}) is given by
$$
\ell^c(\bth,\z,\bpi)=\sum_{i=1}^m\sum_{j=1}^{n_i}\sum_{k=1}^KI(z_{ij}=k)\log \big\{\pi_{ik}h_k(y_{ij}|\x_{ij},\bphi_k)\big\}
+\sum_{i=1}^m\log p(\bpi_i|\bal),
$$
where $p(\bpi_i|\bal)$ denotes the density function of ${\rm Dir}(\bal)$.
Then, given the value of $\bth$ in the $t$th iteration denoted by $\bth^{(t)}$, the E-step entails the imputation of the latent variables $\z$ and $\bpi$ by taking expectation 
$$
Q(\bth|\bth^{(t)})=\E[\ell^c(\bth,\z,\bpi)|Y,\bth^{(t)}],
$$
where the expectation is taken with respect to the posterior distribution of $(\w,\bpi)$ given all the response variables $Y$. 
However, since an analytical form of $Q(\bth|\bth^{(t)})$ is not available, we consider Monte Carlo approximation of $Q(\bth|\bth^{(t)})$ as 
$$
Q(\bth|\bth^{(l)})\approx\frac{1}{L}\sum_{l=1}^L \ell^c(\bth,\z^{(l)},\bpi^{(l)}),
$$
where $L$ is a sufficiently large number, and $\z^{(l)}$ and $\bpi^{(l)}$ are the $l$th random sample generated from the posterior distribution of $(\z,\bpi)$ given $Y$ with $\bth=\bth^{(t)}$. 
Under the hierarchy (\ref{hie}), the marginal posterior distributions of $\z$ and $\bpi$ are not simple forms, but the full conditional distributions of $\z|\bpi,Y$ and $\bpi|\z,Y$ are the following familiar distributions:
\begin{equation}\label{pos}
\begin{split}
&z_{ij}|\bpi_i,Y \sim \text{Cat}(K,\bpt_{ij}),\quad i=1,\dots,m, \quad j=1,\dots,n_i,\\
&\bpi_i|\z, Y \sim \text{Dir}(\bat_i),\quad i=1,\dots,m,
\end{split}
\end{equation}
where $\bpt_{ij}=(\pt_{ij1},\ldots,\pt_{ijK})^t$ and $\bat_i=(\at_{i1},\ldots,\at_{iK})^t$ with
$$
\pt_{ijk}=\frac{\pi_{ik}h_k(y_{ij}|\x_{ij},\bphi_k^{(t)})}{\sum_{l=1}^K\pi_{il}h_l(y_{ij}|\x_{ij},\bphi_l^{(t)})}, \ \ \ \text{and} \ \ \ \ 
\at_{ik}=\al_k^{(t)}+\sum_{j=1}^{n_i}I(z_{ij}=k).
$$
Then we can use a Gibbs sampler for generating random samples of the posterior distribution of $(\z,\bpi)$.

The M-step maximizes $Q(\bth|\bth^{(t)})$ obtained from the E-step, noting that 
$$
Q(\bth|\bth^{(t)})=C+\sum_{i=1}^m\sum_{j=1}^{n_i}\sum_{k=1}^Kz_{ijk}^{\ast}\log h_k(y_{ij}|\x_{ij},\bphi_k)+\sum_{i=1}^m\E[\log p(\bpi_i|\bal)|Y,\bth^{(t)}],
$$ 
where $C$ is a constant independent of $\bth$ and $z_{ijk}^{\ast}=\E[I(z_{ij}=k)|Y,\bth^{(t)}]$ computed from the E-step.
Therefore, the maximization problem of $Q(\bth|\bth^{(t)})$ can be divided into the following: 
\begin{equation}\label{max}
\begin{split}
&\bphih_k=\underset{\bphi_k}{\rm argmax}\sum_{i=1}^m\sum_{j=1}^{n_i}z_{ijk}^{\ast}\log h_k(y_{ij}|\x_{ij},\bphi_k), \ \ \ \ k=1,\ldots,K,\\
&\balh=\underset{\bal}{\rm argmax} \left\{m\log\Gamma\Big(\sum_{k=1}^K\al_k\Big) -m\sum_{k=1}^K\log\Gamma(\al_k) +\sum_{k=1}^K \al_k\sum_{i=1}^m(\log \pi_{ik})^{\ast} \right\},
\end{split}
\end{equation}
where $(\log \pi_{ik})^{\ast}=\E[\log \pi_{ik}|Y,\bth^{(t)}]$.
It is noted that the maximization with respect to each $\bphi_k$ is identical to maximizing the weighted log-likelihood function of the latent conditional distributions, which can be easily carried out by using, for example, the Newton-Raphson algorithm.  
Similarly, the maximization with respect to $\bal$ is similar to performing the maximum likelihood method in the Dirichlet distribution and is not difficult.

The whole procedure of the proposed MCEM algorithm is summarized as follows.

\bigskip
\noindent
\begin{algo}[MCEM algorithm] 
Iterative:
\begin{itemize}
\item[1.]
Set the initial values $\bth^{(0)}$ and $t=0$. 

\item[2.]
Draw a large number of samples $\bpi$ and $\z$ by Gibbs sampling with the full conditionals (\ref{pos}), and compute $z_{ijk}^{\ast}=\E[I(z_{ij}=k)|Y,\bth^{(t)}]$ and $(\log \pi_{ik})^{\ast}=\E[\log \pi_{ik}|Y,\bth^{(t)}]$.

\item[3.]
Solve the maximization problem \eqref{max} and set $\bphi_k^{(t+1)}=\bphih_k$ and $\bal^{(t+1)}=\balh$. 

\item[4.]
If the algorithm has converged, the the algorithm is terminated. 
Otherwise, set $t=t+1$ and go back to Step 2. 
\end{itemize} 
\end{algo}

\bigskip
In the case of the normal linear regression model as the latent model, namely $F_k(\x_{ij},\bphi_k)=N(\x_{ij}^t\bbe_k,\sigma^2_k)$ in (\ref{hie}), the M-step for $\bphi_k=(\bbe_k^t,\sigma_k^2)^t$ in (\ref{max}) can be obtained analytically:
\begin{align*}
&\bbeh_k=\bigg(\sum_{i=1}^m\sum_{j=1}^{n_i} z_{ijk}^{\ast}\x_{ij}\x_{ij}^t\bigg)^{-1}\sum_{i=1}^m\sum_{j=1}^{n_i} z_{ijk}^{\ast}\x_{ij}y_{ij}, \\ 
&\sih^2_k=\bigg(\sum_{i=1}^m\sum_{j=1}^{n_i}  z_{ijk}^{\ast}\bigg)^{-1}\sum_{i=1}^m\sum_{j=1}^{n_i} z_{ijk}^{\ast}(y_{ij}-\x_{ij}^t\bbeh_k)^2. 
\end{align*}
for $k=1,\ldots,K$.

Following Shi and Copas (2002), the convergence of the proposed MCEM algorithm is monitored by using the batch mean $\widetilde{\bth}^{(t)}=H^{-1}\sum_{h=0}^{H-1}\bth^{(t-h)}$, 
after the $H$th iteration.
The algorithm is terminated when the relative difference $\|\widetilde{\bth}^{(t)}-\widetilde{\bth}^{(t-d)}\|/(\|\widetilde{\bth}^{(t-d)}\|+\delta)$, is smaller than some predetermined (small) $\ep$.
Here, $H$, $d$, $\ep$ and $\delta$ are specified by the user, and we use $H=30$, $d=5$, $\ep=\delta=0.001$ as default choices. 
For the E-step, $L=500$ is used as the default choice and this choice appears to work well in the numerical examples in Section~\ref{sec:num}.

For selecting the number of latent distributions, $K$, we use the Akaike information criteria (AIC) or the Bayesian information criteria (BIC) based on the log-marginal likelihood, without any theoretical justifications.
When $\bphi_k$ is $p$-dimensional, the number of parameters included in the model (\ref{hie}) is $pK+K$. 
Then the formulations of AIC and BIC are given by
\begin{align*}
{\rm AIC} &= -2\sum_{i=1}^m \log  f^m_i(\y_{i} | \x_{i},\bthh)  + 2(pK+K), \\
{\rm BIC} &= -2\sum_{i=1}^m \log f^m_i(\y_{i} | \x_{i},\bthh)  + (pK+K)\log N,
\end{align*}
where $N=\sum_{i=1}^m n_i$ is the total number of observations and 
\begin{equation}\label{ml}
f_i^m(\y_{i}|\x_{i},\bthh) = \int \left\{\prod_{j=1}^{n_i} \sum_{k=1}^K \pi_{ik}h_k(y_{ij}|\x_{ij},\bphih_k) \right\}p(\bpi_i|\balh)d\bpi_i
\end{equation}
is the maximum marginal likelihood.
As noted in Section \ref{sec:est}, since the direct evaluation of the marginal likelihood is computationally prohibitive, the maximum marginal likelihood is evaluated by the Monte Carlo integration.
Let $\bpi_i^* = (\pi_{i1}^*,\dots,\pi_{iG}^*)^t$ be the random vector generated from ${\rm Dir}(\balh)$.
Then, the Monte Carlo approximation of (\ref{ml}) is
\begin{align*}
f_i^m(y_{i}|\x_{i},\bthh) \approx \frac{1}{B}\sum_{b=1}^B\left\{\prod_{j=1}^{n_i} \sum_{k=1}^K \pi_{ik}^{*(b)} h_k(y_{ij}|\x_{ij},\bphih_k)\right\},
\end{align*}
for a large $B$, where $(\pi_{i1}^{*(b)},\dots,\pi_{iK}^{*(b)})^t$ is the $b$th draw from ${\rm Dir}(\balh)$.

Let $K^{\ast}$ be the selected number of latent distributions based on AIC or BIC.
Then the feasible version of the cluster-wise estimated conditional density (\ref{Est}) is given by
$$
\widehat{f}_i(y|\x)=\sum_{k=1}^{K^{\ast}}\widehat{\pi}_{ik}h_k(y|\x,\bphih_k),
$$
where $\widehat{\pi}_{ik}=\E[\pi_{ik}|Y_i]$ evaluated at $\bth=\bthh$, which can be computed via the Gibbs sampler (\ref{pos}) with $\bth=\bthh$.


\subsection{Flexible modeling of mixing proportions}\label{sec:CD}
One possible criticism for the formulation of the proposed latent mixture model (\ref{model}) is its simplicity in the relationship between the response variable $y$ and covariate vector $\x$.
In the context of mixture modeling for non-clustered (independent) data, Geweke and Keane (2007) proposed a flexible modeling of the mixing proportions by considering covariate dependent structures.
Then, we here consider implementing the idea to the modeling cluster-wise conditional densities, that is, we consider the following structure in the distribution of the mixing proportions:
\begin{equation}\label{CD}
\bpi_i\sim {\rm Dir}(\bal_i), \ \ \ \ \bal_i=(\al_{i1},\ldots,\al_{iK})^t, \ \ \ \
\al_{ik}=\exp(\w_i^t\bga_k),
\end{equation}
where $\w_i$ is the $q$-dimensional vector of the cluster-specific covariates and $\bga_k$ is the corresponding coefficient. 
One can take, for example, $\w_i=\barx_i^{(s)}$ where $\barx_i^{(s)}=n_i^{-1}\sum_{j=1}^{n_i} \x_{ij}^{(s)}$ and $\x_{ij}^{(s)}$ is the subvector of $\x_{ij}$.
Under this setting, it hods that 
$$
E[\pi_{ik}]=\frac{\exp(\w_i^t\bga_k)}{\sum_{k=1}^K\exp(\w_i^t\bga_k)}.
$$
the MCEM algorithm developed in Section \ref{sec:est} can be easily modified to estimate the model with  (\ref{CD}). 
Specifically, in the E-step $\at_{ik}$ appeared in the full conditional distribution of $\bpi_i|\w,Y$ in (\ref{pos}) is replaced with 
$$
\at_{ik}=\exp(\w_i^t\bga^{(t)}_k)+\sum_{j=1}^{n_i}I(z_{ij}=k),
$$
and the M-step for $\bal$ in (\ref{max}) is replaced with the maximizing 
\begin{align*}
Q(\bga)
&=\sum_{i=1}^m\log\Gamma\Big\{\sum_{k=1}^K\exp(\w_i^t\bga_k)\Big\}
-\sum_{i=1}^m\sum_{k=1}^K\log\Gamma(\exp(\w_i^t\bga_k)) \\
& \ \ \ \ \ +\sum_{i=1}^m\sum_{k=1}^K \exp(\w_i^t\bga_k)(\log \pi_{ik})^{\ast},
\end{align*}
where $\bga=\{\bga_1,\ldots,\bga_K\}$.
Finally, it is noted that the number of parameters under (\ref{CD}) is $K(p+q)$, so that the penalty terms in AIC and BIC used for selecting $K$ should be changed accordingly.

\section{Numerical Studies}\label{sec:num}


\subsection{Simulation studies}
The finite sample performance of the proposed latent mixture model is investigated together with some existing methods.
We consider two cases of within-cluster sample sizes $n_i=30$ and $n_i=50$ for $i=1,\dots,m$ and $m=50$. 
For the true conditional density in the $i$th cluster, the following two scenarios are considered:
\begin{align*}
\text{(I)} \ \ \ \ &f_i(y|x)=\pi_i\phi(y; -1+x, 1)+(1-\pi_i)\phi(y; 1-x, 1), \ \ \ \ \pi_{i}\sim \text{Beta}\left(5,3\right),\\
\text{(II)} \ \ \ \ &f_i(y|x)=I(1\leq i\leq 15)\phi(y; -1+2x, 0.5^2)+I(16\leq i\leq 30)\phi(y; 1.5+x, 1)\\
& \ \ \ \ \ \ \ \ \ \ \ \ \ \ \ \ +I(31\leq i\leq 50)\phi(y;  -x, 1.5^2),
\end{align*}
where $i=1,\ldots,m$, and $\phi(\cdot; a,b)$ denotes the density function of the normal distribution $N(a,b)$ and $x_{ij}\sim N(0,1)$ in each scenario. 
The latent mixture regression (LMR)  model with normal linear regression models used as latent models is considered, and the number of latent components are selected by using BIC. 
For comparison, we also consider the local mixture (LM) model where the mixture of normal linear regressions is fitted to each cluster separately and global mixture (GM) model where the single mixture of normal linear regressions is fitted to the whole data ignoring the cluster heterogeneity. 
For both models, the number of components was selected based on BIC. 
Moreover, as the competitor from random effect models, we also applied a random intercept (RI) model.
Note that GM ignores the clustering structure and produces the same conditional densities over all the clusters.
On the other hand, while LM may flexibly express the cluster-wise conditional density, the results are expected to be unstable due to the relatively small within-cluster sample sizes.

The performance of the models is measured based on the cluster-wise mean integrated squared error (MISE) defined as 
$$
\text{MISE}_i=\frac1R\sum_{r=1}^R\int \left\{\widehat{f}^{(r)}_i(t|x)-f_i(t|x)\right\}^2 \text{d}t,\quad i=1,\dots,m,
$$
where $\widehat{f}^{(r)}_i(t|x)$ is the estimated conditional density obtained from the $r$th replication. 
Since the above MISE depends on the covariate $x$, we considered the three values, $x=-1.5,-0.75, 0$. 
We computed the cluster-wise MISE of four models based on $R=100$ replications. 

\begin{figure}[!htb]
\centering
\includegraphics[width=12cm]{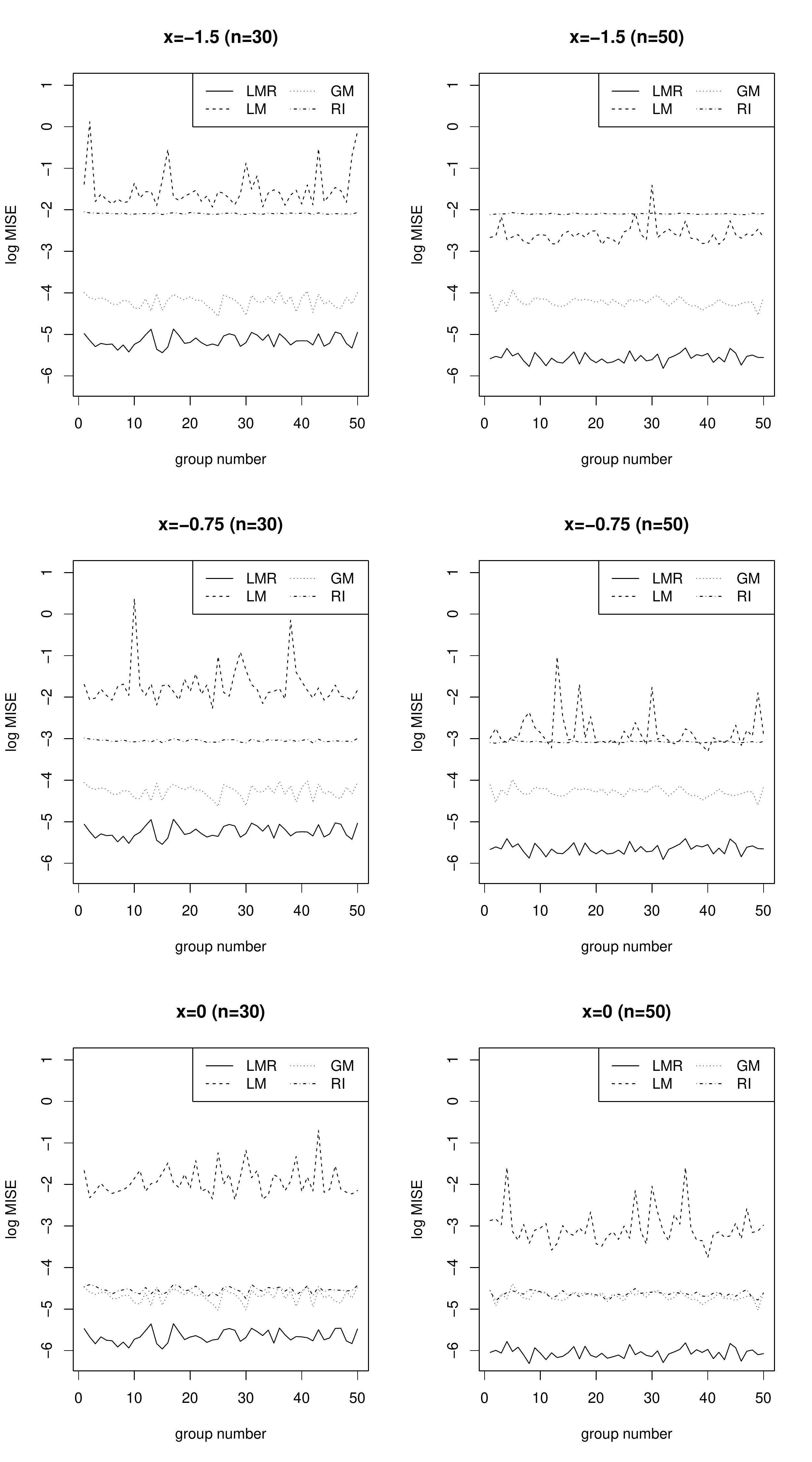}
\caption{Mean integrated squared error (MISE) of four models evaluated at $x=-1.5,-0.75, 0$ in scenario (I) with $n=30$ (left) and $n=50$ (right).
\label{fig:s1}
}
\end{figure}

\begin{figure}[!htb]
\centering
\includegraphics[width=12cm]{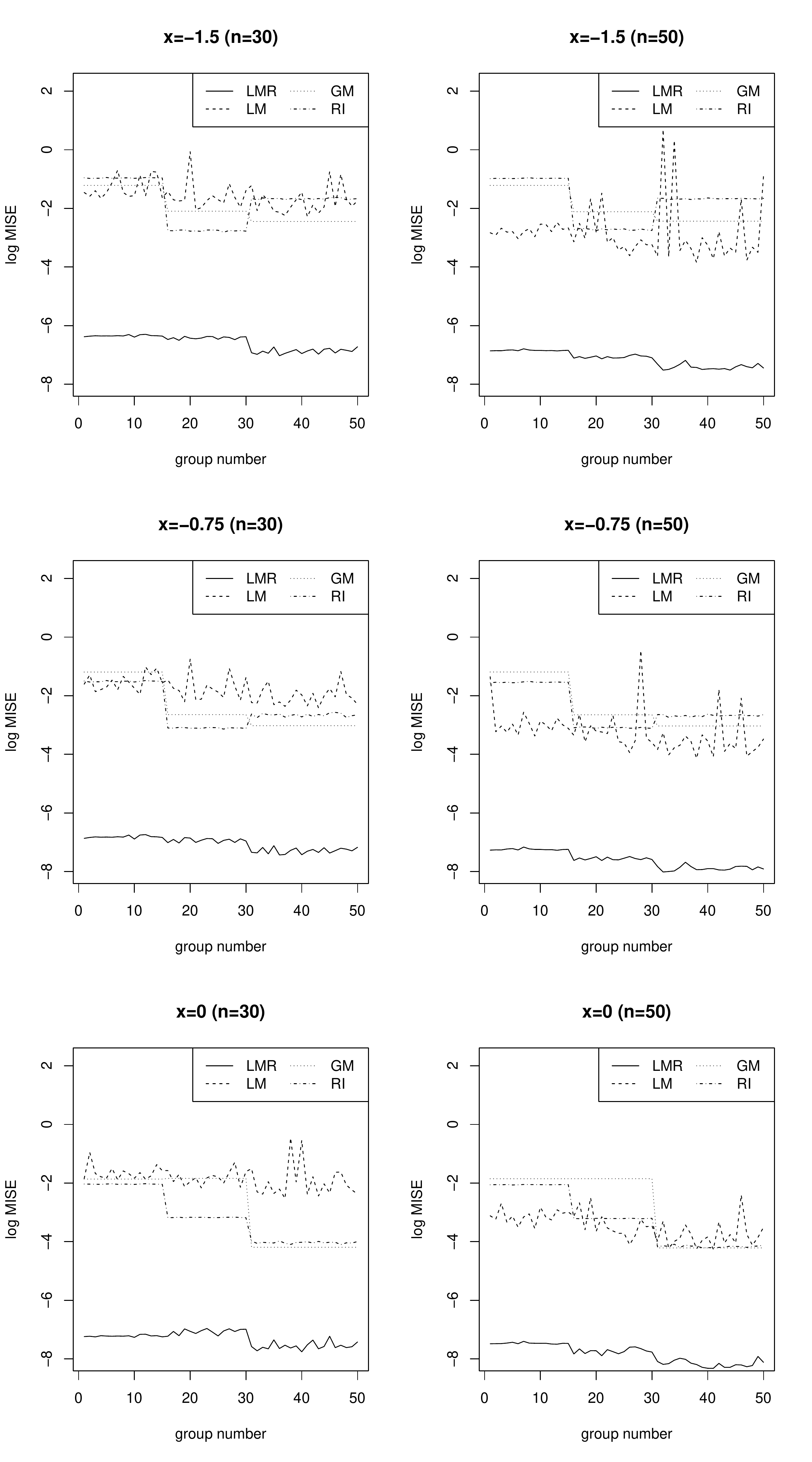}
\caption{Mean integrated squared error (MISE) of four models evaluated at $x=-1.5,-0.75, 0$ in scenario (II) with $n=30$ (left) and $n=50$ (right).
\label{fig:s2}
}
\end{figure}

Figure~\ref{fig:s1} and \ref{fig:s2} present the cluster-wise MISE for Scenario (I) and (II), respectively. 
The figures show that the proposed LMR model outperforms in all cases.
As expected, LM appears to have produced the unstable results due to the relatively small sample sizes in spite of its flexibility.
On the other hand, GM seems to perform relatively well in this study as the number of parameters is small compared with LM.
However, since GM produces the same conditional density estimators over the clusters, GM performs no better than LMR. 
Concerning RC, it may perform as well as GM for $x=0$ in Scenario~(I) and some cases in Scenario~(II), but the result is much inferior to that of LMR. 
Although not shown here, BIC selected the true number of components most of the time, while the selected number of components by AIC tended to be larger than the truth. 
Hence, BIC would be preferable to AIC and only the results based on BIC are considered in the rest of this article.

\begin{figure}[!htb]
\centering
\includegraphics[width=12cm]{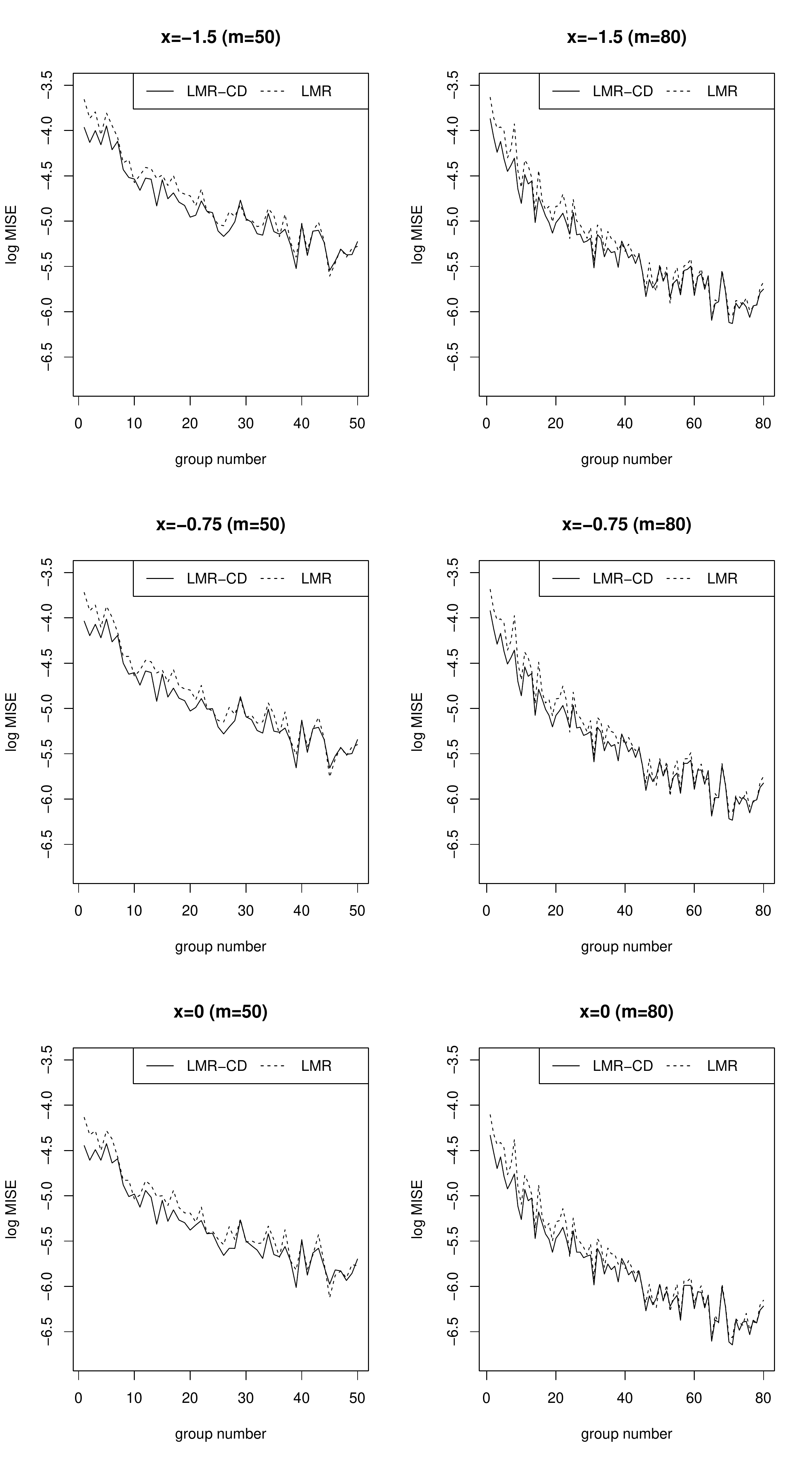}
\caption{Mean integrated squared error (MISE) of three models evaluated at $x=-1.5,-0.75, 0$ in scenario (III) with $m=50$ (left) and $m=80$ (right).
\label{fig:s3}
}

\end{figure}

We next investigate the efficacy of the modeling the distribution of the mixing proportion in terms of some covariates as introduced in Section \ref{sec:CD}.
To this end, we consider the following true conditional density:
\begin{align*}
\text{(III)} \ \ \ \ \ &f_i(y|x)=\pi_i \phi(y; -1+x,1)+(1-\pi_i)\phi(y; 1-x,1), \ \ \ \ \pi_i\sim \text{Beta}(\alpha_{i1},\alpha_{i1}),\\
&\ \ \  \alpha_{i1}=\exp(1+0.6w_i), \ \ \ \ \alpha_{i2}=\exp(1-0.5w_i), \ \ \ w_i\sim \text{Ber}(0.4).
\end{align*}
We set $n_i=i$ for $i=1,\dots,m$ such that the within-cluster sample size varies across clusters and  consider two cases of $m$, $m=50$ and $80$.
As in the previous studies, the covariates $x_{ij}$'s are generated from $N(0,1)$.
The latent mixture regression model with covariate-dependent structure of mixing proportions (LMR-CD) and the latent mixture regression model (LMR) are fitted to the simulated data.
For both models, we use the normal linear regression models as the component models, and the number of components is selected based on BIC.
For comparison, we again computed the MISE with $x=-1.5,-0.75, 0$, and the results are presented in Figure \ref{fig:s3}.
In the figure, LMR-CD appears to perform better than LMR for the clusters with the small within-cluster sample sizes for both $m$.

\subsection{Real data example}
\begin{figure}[!htb]
\centering
\includegraphics[width=14cm]{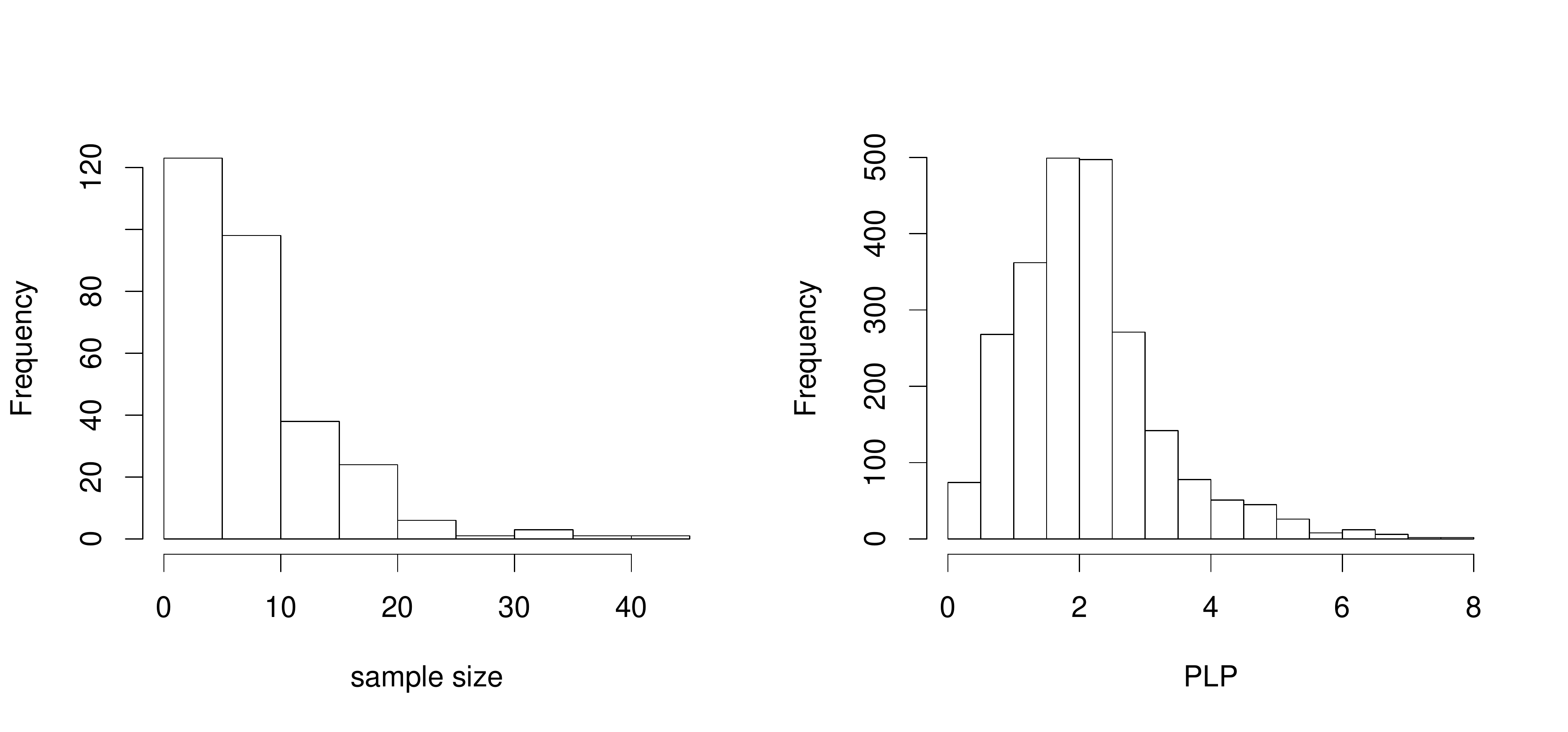}
\caption{Histograms of within-cluster sample size $n_i$ (left) and posted land price $y_{ij}$ (right).
\label{fig:PLP1}
}
\end{figure}

To demonstrate the proposed method in a practical situation, we apply the latent mixture model to the posted land price (PLP) data in Tokyo and the surrounding four prefectures (Chiba, Saitama, Kanagawa and Ibaraki) in 2001.
The data units (locations) are clustered with respect to the nearest station.
The number of clusters is $m=295$ and the total number of units is $N=2363$.
The number of within-cluster samples $n_i$ are ranging from $1$ to $45$, and the histogram of $n_i$ is provided in the left panel in Figure \ref{fig:PLP1}.
We note that there are $221$ clusters with $n_i$ smaller than $10$ and $25$ clusters with $n_i=1$.
The response variable $y_{ij}$ is the PLP which is measured in 100,000 yen per squared meter. 
In each $j$th unit (location) in $i$th cluster (station), $y_{ij}$ is observed with the floor area ratio (\%) $F_{ij}$ and amount of time $A_{ij}$ (second) to station $i$ on foot.
Moreover, as cluster level information, the amount of time $T_i$ from Tokyo station by train and the prefecture to which the station belongs are available.
We use four dummy variables $D_{i1},D_{i2},D_{i3}$, and $D_{i4}$ for Chiba, Saitama, Kanagawa, and Ibaraki, respectively, which take value one if the station $i$ belongs to the corresponding prefecture and zero otherwise. 
The values of $y_{ij}$ range from $0.158$ to $20.3$.
The right panel of Figure \ref{fig:PLP1} shows that the histogram of $y_{ij}$ for $y_{ij}<8$. 
Note that the number of samples with $y_{ij}\geq8$ is only $20$ which is less than $1\%$ of the total number of observations. 
Using this dataset, the conditional density of the PLP for each station is estimated.

Let $\x_{ij}=(1,F_{ij},A_{ij},T_i,D_{i1},\ldots, D_{i4})^t$.
We consider the following latent mixture regression (LMR) model:
\begin{equation}\label{PLP-LMR}
\begin{split}
&f_i(y_{ij}|\pi_{i1},\ldots,\pi_{iK})=\sum_{k=1}^K\pi_{ik}\phi(y_{ij};\x_{ij}^t\bbe_k,\si_k^2),  \ \ \ j=1,\ldots,n_i, \ \ \ i=1,\ldots,m, \\
&(\pi_{i1},\ldots,\pi_{iK})^t\sim \text{Dir}(\alpha_{i1},\ldots,\alpha_{iK}), \ \ \ \ \ 
\alpha_{ik}=\exp(\gamma_{1k}+\gamma_{2k}T_i^{\ast}), \ \ \ k=1,\ldots,K,
\end{split}
\end{equation}
where $\phi(\cdot;a,b)$ denotes the density function of $N(a,b)$, and $T_i^{\ast}$ is the standardized version of $T_i$.
It is noted that the marginal model (\ref{mar}) is given by
\begin{equation}\label{PLP-mar}
f_i(y_{ij})=\sum_{k=1}^K p_{ik}\phi(y_{ij};\x^t_{ij}\bbe_k,\si_k^2), \ \ \ \ \ 
p_{ik}=\frac{\alpha_{ik}}{\sum_{\ell=1}^K\alpha_{i\ell}},
\end{equation}
and the cluster-wise estimated density (\ref{Est}) is
$$
f_i(y)=\sum_{k=1}^K \E[\pi_{ik}|Y_i]\phi(y;\x^t\bbe_k,\si_k^2), 
$$
where $Y_i=\{y_{i1},\ldots,y_{in_i}\}$ and $\E[\pi_{ik}|Y_i]$ can be computed from the Gibbs sampling (\ref{pos}).
Moreover, based on BIC, the number of latent components was selected to be $K=6$ from $\{1,\dots, 8\}$.
We also doubled  the number of  Gibbs draws in the E-step, but the same result was obtained. 

For comparison with the proposed method, we also applied the global mixture (GM) model with $K_{\ast}$ components:
$$
f(y)=\sum_{k=1}^{K_{\ast}}p_k\phi(y;\x^t\bbe_k,\si_k^2), 
$$
where $\sum_{k=1}^{K_{\ast}}p_k=1$.
It is expected that the estimated GM is similar to the marginal model in LMR.
Based on BIC $K_{\ast}=5$ was selected.

\begin{figure}[!htb]
\centering
\includegraphics[width=14cm]{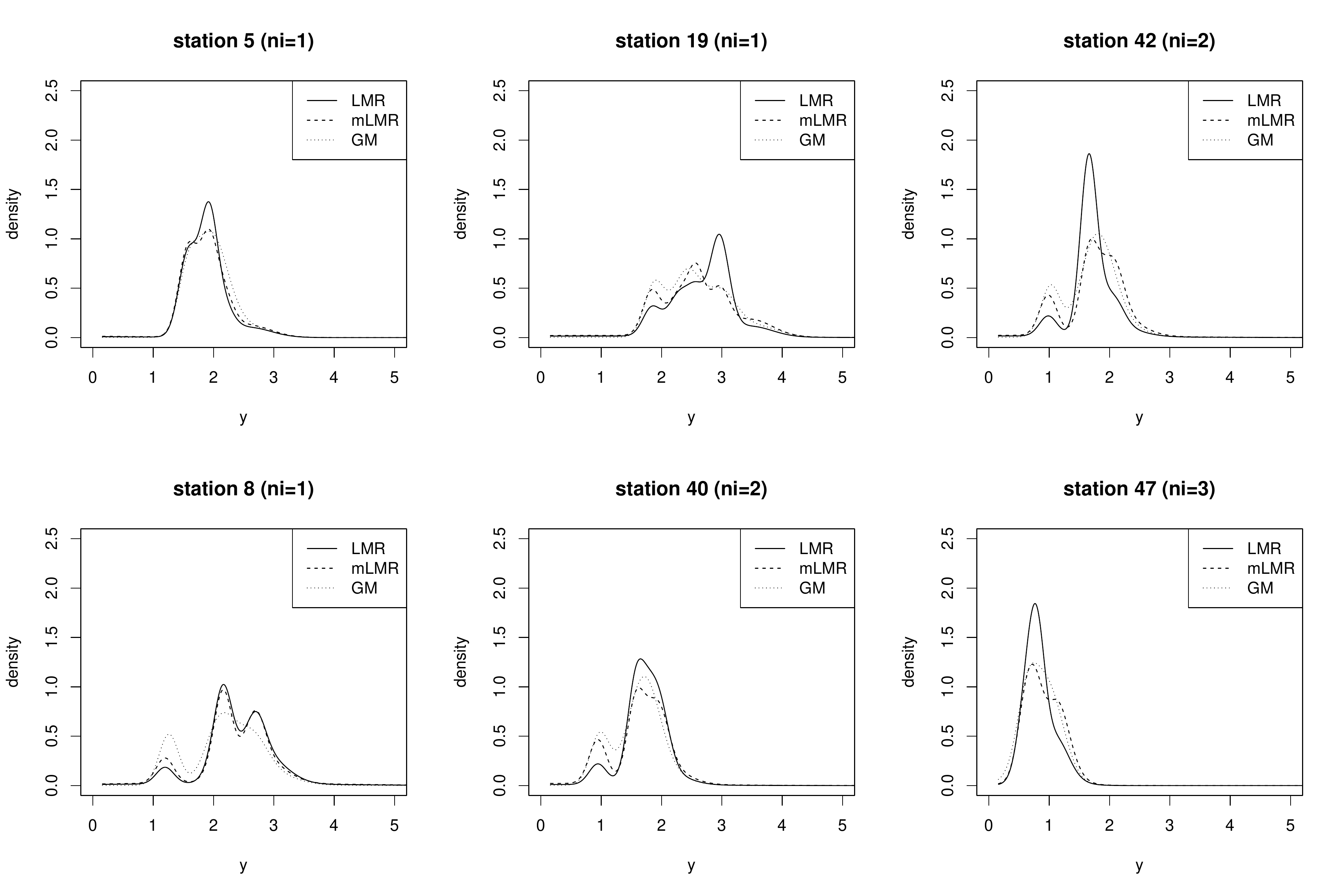}
\caption{Estimated cluster-wise conditional densities of PLP in stations with small $n_i$.
\label{fig:PLP-est1}
}
\end{figure}

\begin{figure}[!htb]
\centering
\includegraphics[width=14cm]{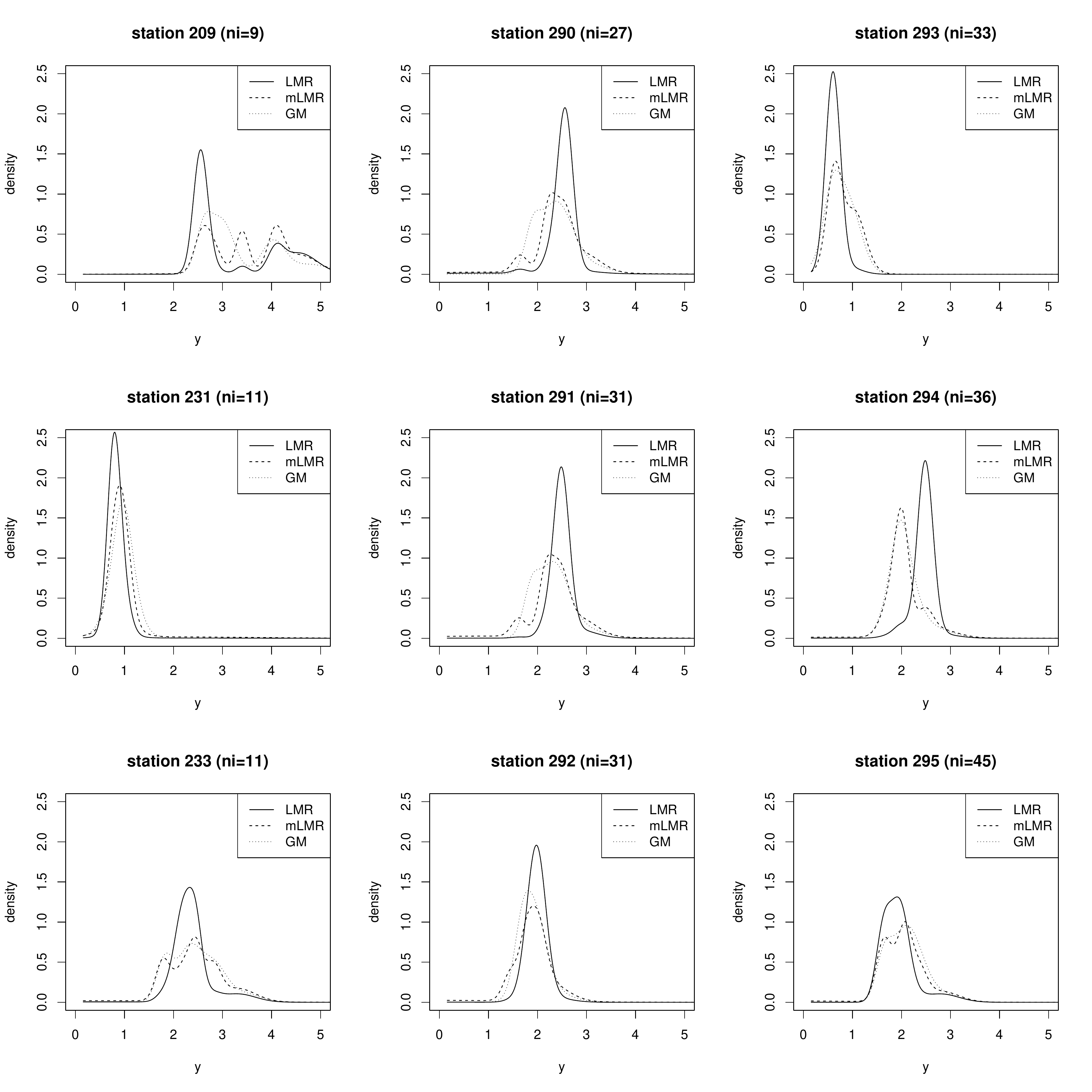}
\caption{Estimated cluster-wise conditional densities of PLP in stations with moderate or large $n_i$.
\label{fig:PLP-est2}
}
\end{figure}

To visualize the estimated conditional density in each cluster, we fixed the covariate vector $\x$ at $(1,100,600,T_i,D_{i1},\ldots, D_{i4})^t$, in which $f_i(y|x)$ corresponds to the density function of the PLP of each cluster when the floor area ratio is $100$ and the location is $10$ minutes' walk from the nearest station.  
Figure \ref{fig:PLP-est1} presents the estimated density under LMR, the marginal model of LMR (mLMR), and GM for the stations with small $n_i$. 
The figure shows that the cluster-wise estimated densities under LMR are close to those under the marginal model (\ref{PLP-mar}) when $n_i$ is small.
This is because the small $n_i$ values leads to a small difference between the prior mean $p_{ik}$ and posterior mean $\E[\pi_{ik}|Y_i]$ of $\pi_{ik}$, so that the estimated densities in such clusters are automatically close to those under the marginal model which can be stably estimated from the data.
Figure \ref{fig:PLP-est2} presents the estimated densities for the stations with relatively large $n_i$. 
Contrary to Figure \ref{fig:PLP-est1}, the estimated densities under LMR are apart from the marginal model in some clusters.
The result implies that the marginal model is adjusted by the observed data in these clusters.
We finally point out that the marginal model of LMR and GM are similar in most cases since their modeling strategies are similar in the sense that they aim at estimating the global density by ignoring the clustering structure.

\vspace{20cm}

\section{Conclusion and Discussion}\label{sec:diss}
We have proposed the latent mixture model for estimating the cluster-wise conditional distributions.
The model parameters are estimated by using the simple Monte Carlo EM algorithm instead of the brute force maximization of the marginal likelihood. 
Through the simulation and empirical studies, the proposed method is found to be useful for flexible modeling of clustered data.

In this article, we selected the number of components by using AIC and BIC.
However, it is well-recognized that the mixture model is a singular model and the use of AIC or BIC is not justified.
The detailed investigation of selecting the number of latent components with theoretical validity would extend the scope of this article, which will be left as a valuable future work.

\medskip
\bigskip
\noindent
{\bf Acknowledgments.}\ \ \ \
This work was supported by JSPS KAKENHI Grant Numbers [16H07406, 15K17036, 16K17101].
The computational results were obtained using Ox version~6.21 (Doornik, 2007).

\medskip

\end{document}